\documentclass[aps,prl,reprint,superscriptaddress]{revtex4-1}
\usepackage{braket}
\usepackage{graphicx}
\usepackage{amsmath}
\usepackage{amssymb}
\usepackage{comment}
\usepackage{bbold}
\usepackage[usenames]{color}

\renewcommand{\vec}[1]{\mathbf{#1}}
\usepackage{bm}

\begin{document}

\title{ Anomalous Nonlocal Resistance and Spin-charge Conversion Mechanisms in Two-Dimensional Metals}

\author{Chunli Huang}
\affiliation{
 Division of Physics and Applied Physics, School of Physical and Mathematical Sciences,
Nanyang Technological University, Singapore 637371, Singapore
}
\affiliation{Department of Physics, National Tsing Hua University, Hsinchu 30013, Taiwan}

\author{Y. D. Chong}
\affiliation{
 Division of Physics and Applied Physics, School of Physical and Mathematical Sciences,
Nanyang Technological University, Singapore 637371, Singapore
}

\author{Miguel A. Cazalilla}
\affiliation{Department of Physics, National Tsing Hua University and National Center for Theoretical Sciences (NCTS), Hsinchu 30013, Taiwan}

\date{\today}

\begin{abstract}
We uncover two anomalous features in the nonlocal transport behavior of two-dimensional metallic materials with spin-orbit coupling. Firstly, the nonlocal resistance can have negative values and oscillate with distance, even in the absence of a magnetic field.  Secondly, the oscillations of the nonlocal resistance under an applied in-plane magnetic field (Hanle effect) can be asymmetric under field reversal.  Both features are produced by direct magnetoelectric coupling, which is
possible in materials with broken inversion symmetry but was not included in previous spin diffusion theories of nonlocal transport.  These effects can be used to identify the relative contributions of different spin-charge conversion mechanisms. They should be observable in adatom-functionalized graphene, and may provide the reason for discrepancies in recent nonlocal transport experiments on graphene.
\end{abstract}

\maketitle

\textit{Introduction}---The ability to convert between macroscopic spin and charge degrees of freedom is a distinctive feature of materials with spin-orbit coupling (SOC), and has fundamental importance for spintronics research~\cite{RevModPhys.76.323}.  The two most prominent examples of spin-charge conversion are the spin Hall effect (SHE)~\cite{Dyakonov_Perel_2,sinova2015spin} and current-induced spin polarization (CISP)~\cite{Edelstein1990233,rashba1984}: when an electric current $\vec{J}$ is injected into a material with SOC, it can generate a spin current $\boldsymbol{\mathcal{J}}$ (the SHE), and/or a non-equilibrium spin polarization $\vec{m}$ (CISP).  Spin-charge conversion can be detected and studied using nonlocal transport experiments \cite{abanin2009nonlocal,seki2008giant,balakrishnan2014giant,kaverzin2015electron}, a well-established technique that has been applied to two-dimensional (2D) quantum spin Hall insulators~\cite{Konig766}, 3D topological Kondo insulators~\cite{Fisk_nonlocal}, and many other systems~\cite{sui2015gate,levitov2016electron,PhysRevLett.115.057206,PhysRevX.4.031035}.  
These experiments rely on a combination of spin-charge conversion processes: when $\vec{J}$ is injected at one position, the SHE (CISP) converts part of it to $\boldsymbol{\mathcal{J}}$ ($\vec{m}$), which diffuses across the device, and is then converted back into $\vec{J}$ by the inverse SHE (inverse CISP) and measured via the nonlocal electrical resistance $R_{\mathrm{nl}}$.  Moreover, applying an in-plane magnetic field induces Hanle precession, which is observed as an oscillation of $R_{\mathrm{nl}}$ with distance and field strength \cite{Tombros2007}.

To date, the analysis of spin-charge conversion in nonlocal transport experiments has relied heavily on a theory developed by Abanin~\textit{et~al.}~\cite{abanin2009nonlocal}, which assumes that SHE is the dominant spin-charge conversion mechanism present.
 However, many materials of interest in spintronics have large CISP effects~\cite{manchon2015new,soumyanarayanan2016emergent} arising from Rashba SOC. This is especially so in (quasi) 2D materials with broken inversion symmetry, such as the surfaces states of 3D topological insulators \cite{hailong_surface,fert_zhang_2016}, gold-hybridized graphene \cite{marchenko2012giant,eoin_rashba2016} and Bi/Ag quantum wells \cite{sanchez2013spin}. Recently, a great deal of effort has been put into nonlocal transport experiments on adatom-functionalized graphene \cite{balakrishnan_colossal,balakrishnan2014giant,kaverzin2015electron,neutral2015wang}, which is predicted to exhibit strong Rashba SOC \cite{netoguinea09,weeks2011engineering,stabilizing2012hua,fabian2013spin,fabian2015spin}. The results of these experiments appear to be inconsistent with each other, and with the existing spin-diffusion theory~\cite{abanin2009nonlocal}. 




In this Letter, we present a theoretical analysis of diffusive spin
transport in 2D metals that fully accounts for SHE and CISP processes.
We predict that a previously-neglected ``direct magnetoelectric
coupling'' (DMC) process---a direct coupling between the local current
density $\vec{J}$ and the local spin polarization $\vec{m}$---can
produce nonlocal transport behaviors qualitatively different from
the previous model~\cite{abanin2009nonlocal}.  We point out two specific
features that are experimentally accessible.  Firstly, the nonlocal
resistance $R_{\mathrm{nl}}$ can be negative (i.e., having the
opposite sign from the local resistance $R_{xx}$), even in the absence
of an applied magnetic field.  By contrast, in previous models without
DMC, $R_{\mathrm{nl}}$ is always
positive~\cite{abanin2009nonlocal}.
The second unusual feature is an asymmetry in Hanle
precession with respect to the direction of the in-plane magnetic
field $\vec{B}_{\parallel}$.  When the SHE is dominant, the spins are polarized perpendicular to the 2D material, and the Hanle
precession curve is always symmetrical under reversal of
$\vec{B}_{\parallel}$.  If DMC is sufficiently strong, however, the
Hanle precession curve becomes asymmetrical.  
From this asymmetry and the sign of the nonlocal resistance, it
is possible to determine the relative contributions of different
spin-charge conversion mechanisms in a material.  These anomalous
features may be helpful for guiding future studies of SOC in 2D
metals.

DMC is known to be generically possible in materials with broken
inversion (up-down) symmetry~\cite{levitov1985magnetoelectric,footnote1}.
The usual Rashba-Edelstein CISP effect \cite{Edelstein1990233} is \emph{not} a form of DMC, since it arises from an \textit{indirect} coupling of $\mathbf{J}$ and $\mathbf{m}$,
mediated by the spin current $\boldsymbol{\mathcal{J}}$ \cite{Kashen2014micro,chunli2016direct}.  One microscopic mechanism that can lead to DMC, called ``anisotropic spin-precession scattering'', was recently found by the present authors, in the context of adatom-functionalized graphene~\cite{chunli2016direct}, and will be used in our numerical examples.  This form of DMC arises from quantum interference between different components of SOC impurity potentials~\cite{chunli2016direct}.


\begin{figure}[t]
\includegraphics[width=0.45\textwidth]{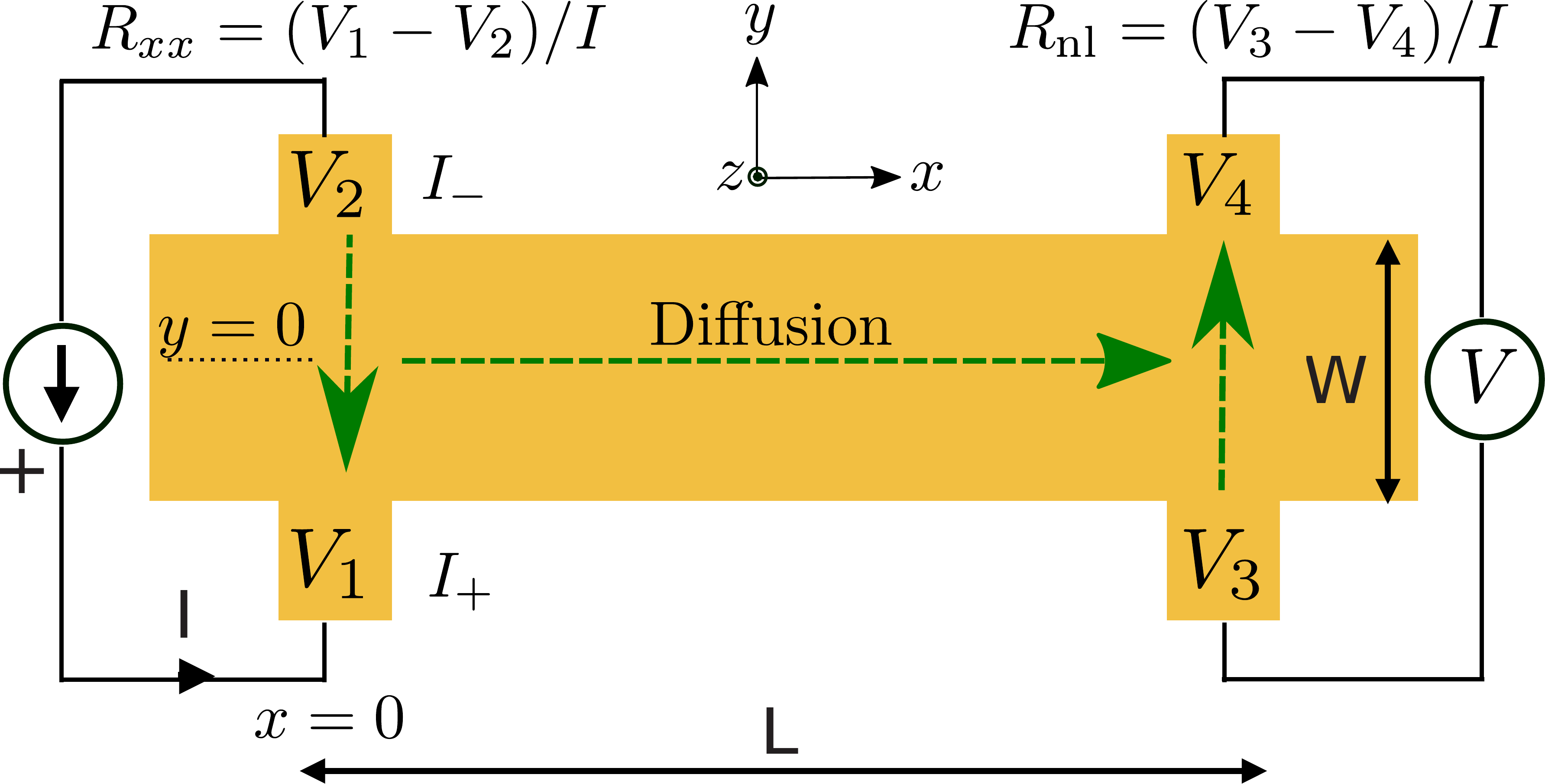}
\caption{Nonlocal transport on a  Hall bar device. A current $I$ is injected at $x=0$ between the electrodes $V_2$ and $V_1$. This gives rise to both a local resistance $R_{xx}$ and a nonlocal resistance $R_{\mathrm{nl}}$. The latter is enhanced by a process involving (i) a spin-charge conversion (current-induced spin polarization or spin Hall effect) at $x=0$, (ii) diffusion from $x=0$ to $x=L$, and (iii) the inverse spin-charge conversion at $x=L$. The green-dotted line indicates the direction of electron flow, following the convention of Ref.~\cite{mihajilov2009prl}.
}\label{fig:scheme}
\end{figure}

\textit{Transport Theory---} We seek to describe the transport of charge and spin in a 2D metal within the diffusive regime ($k_F \ell \gg 1$, where $k_F$ is the Fermi momentum and $\ell$ is the mean free path).  
We start with the spin continuity equation,
\begin{equation}
[\nabla_{t} \,m(\vec{r},t)]^{a}+ [\nabla_{i}{\mathcal{J}}_i(\vec{r},t)] ^a 
= -\frac{m^a(\vec{r},t)}{\tau_{s}} + \kappa^{a}_{j} J_j(\vec{r},t). \label{eq:eom_den}
\end{equation}
Here, lower (upper) indices stand for orbital (spin) components of the current, with orbital coordinates lying in the $x$-$y$ plane, and Einstein's summation convention is used; $\tau_{s}$ is the spin relaxation time (assumed to be isotropic); and $\kappa^{a}_{j}$ parameterizes the DMC, which is a direct local coupling between the magnetization $\mathbf{m}$ and electric current $\mathbf{J}$.  This DMC term was not accounted for in previous theories~\cite{abanin2009nonlocal,raimondi2012su2,shen2014theory} and we shall see it leads to   nontrivial consequences.

 
 
The $\nabla_i$ and $\nabla_{t}$ symbols in Eq.~\eqref{eq:eom_den} are covariant derivatives that account for spin precession induced by SOC and magnetic fields~\cite{raimondi2012su2,shen2014theory}. For any vector $\mathbf{V}$,
\begin{align}  \label{covariant_derivative}
  [\nabla _ i V] ^a &= \partial_i V ^a -\epsilon^{abc}\mathcal{A}_{i}^b V^c \\
 [\nabla _ t V ]^a &= \partial_t V ^a +\omega_{L} \epsilon^{abc}\mathrm{\hat{B}}^b V^c,
\end{align}
where $\partial_i$ ($\partial_t$) denotes a spatial (time) derivative, and $\omega_{L}=g\mu_{B}|\vec{\mathbf{B}}|/\hbar$ is the Larmor precession frequency induced by the magnetic field $\vec{B}$.  $\mathcal{A}$ is a non-Abelian gauge field describing the precession of $\vec{m}$ due to $\boldsymbol{\mathcal{J}}$ [see Eqs.~\eqref{eq:eom_den} and \eqref{covariant_derivative}], and vice versa~\cite{raimondi2012su2,shen2014theory}.  It can arise from SOC processes that are intrinsic (e.g., band structure effects), or extrinsic (e.g., spatially averaged SOC impurities~\cite{chunli2016direct}).  For example, in a 2D electron gas, $\mathcal{A}$ can be extracted from the effective SOC Hamiltonian $H =\frac{1}{2m} \mathcal{A}_{j}^{a} p_{j}\sigma^{a}$, where $m$, $p$ and $s$ are the mass, momentum and spin respectively \cite{raimondi2012su2,shen2014theory}. 


\begin{figure*}
\includegraphics[width=0.95\textwidth]{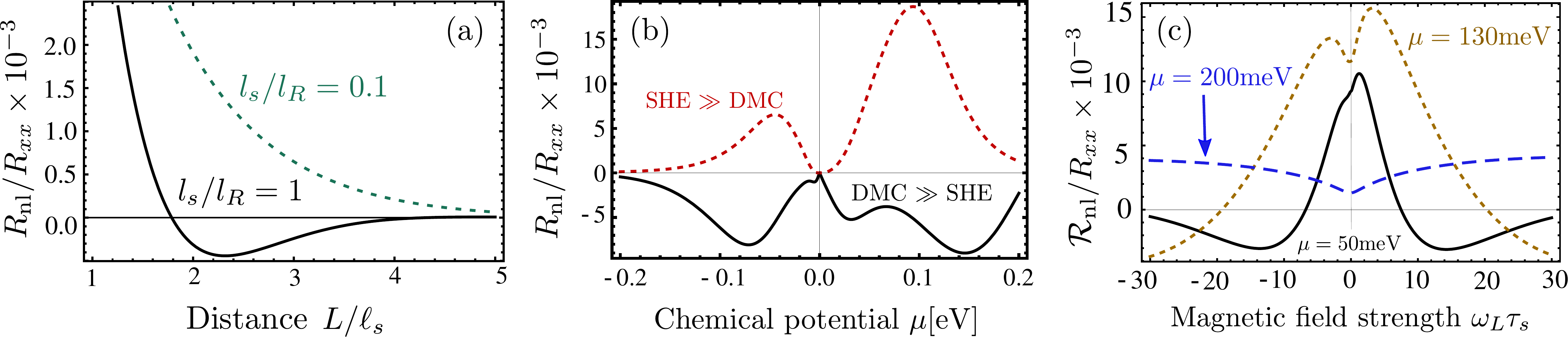}
  \caption{\label{fig:Rnl}
Nonlocal resistance of graphene decorated with adatoms that induce spin-orbit coupling (SOC)~\cite{chunli2016direct}. (a) Nonlocal resistance $R_{\mathrm{nl}}$ versus distance, in the absence of a magnetic field.  When the SOC magnetic length $\ell_{R}$ (i.e. the length scale associated with Rashba SOC) becomes comparable to the spin relaxation length $\ell_s$, $R_{\mathrm{nl}}$ can become negative due to spin precession, even in the absence of a magnetic field. (b) $R_{\mathrm{nl}}$ versus the chemical potential $\mu$ in the absence of a magnetic field, showing that $R_{\mathrm{nl}} > 0$ when the SHE dominates over DMC effects, whereas $R_{\mathrm{nl}} <0$ when DMC dominates.  (c) Hanle precession at different values of $\mu$, in the regime where SHE dominates over DMC \cite{balakrishnan2014giant}. The oscillations are suppressed for values of $\mu$ away from the Dirac point, due to increases in the elastic scattering time.  In all plots, $R_{\mathrm{nl}}$ and $\mathcal{R}_{\mathrm{nl}}$ are normalized to the local resistivity $R_{xx}$.  See Supplemental Materials for details of the microscopic model, such as the SOC potential strengths which determine the conversion factors $\theta_{\mathrm{sH}}$, $\ell_{\mathrm{DMC}}$\cite{SM}.
}
\end{figure*}

To describe the diffusion of spin and charge, Eq.~\eqref{eq:eom_den} must be supplemented by a set of constitutive relations:
\begin{align}
J_i(\vec{r},t) =& -D \, \partial_{i} \rho(\vec{r},t) + \sigma E_i  \notag \\
& - \gamma_{ij}^a \, \mathcal{J}_{j}^{a}(\vec{r},t) +\tau_{c} \kappa^{a}_{i} m^a(\vec{r},t)
\label{eq:eom_c}\\
\mathcal{J}_{i}^{a}(\vec{r},t) =& -D \, [\nabla_{i} m(\vec{r},t)]^a +   \gamma_{ij}^a J_j(\vec{r},t). \label{eq:eom_sc}
\end{align}
Here,  $\gamma_{ij}^a=\theta_{\mathrm{sH}}\epsilon_{ij }\delta^{az}$, where $\theta_{\mathrm{sH}}$ is the spin Hall angle that parameterize the coupling between $\vec{J}$ and $\boldsymbol{\mathcal{J}}$.
$E_i$ is the applied electric field, $\sigma$ is the charge conductivity, $\tau_{c}$ is the elastic charge scattering time,  $D=1/2 v_{F}^{2} \tau_{c}$ is the charge and spin diffusion constant (for simplicity, we assume that charge and spin diffusion are isotropic and share the same diffusion constant), and $\rho(\mathbf{r},t)$ is the charge density. The latter obeys the conservation equation $\partial_{t} \rho + \partial_{i} J_{i}=0$. Note that Eq.~\eqref{eq:eom_sc} uses the covariant derivative defined in Eq.~\eqref{covariant_derivative}.  Moreover, Eq.~\eqref{eq:eom_c} contains a DMC term, with $\kappa_{i}^a$ entering with the same sign as in Eq.~\eqref{eq:eom_den}, consistent with Onsager's reciprocity principle.  \cite{giuliani_giovanni}. (In Eqs.~\eqref{eq:eom_c} and \eqref{eq:eom_sc}, $\gamma_{ij}^a$ enters with opposite signs, again consistent with Onsager reciprocity.)

Eqs.~\eqref{eq:eom_den}--\eqref{eq:eom_sc} contain two distinct processes that contribute to CISP.  The first is the Rashba-Edelstein effect \cite{Edelstein1990233,Kashen2014micro}: a charge current $\vec{J}$ induces a spin current $\boldsymbol{\mathcal{J}}$ via the SHE, [i.e.~the $\gamma_{ij}^a$ term in Eq.~\eqref{eq:eom_c}], then $\boldsymbol{\mathcal{J}}$ precesses (or  induces) the spin density $\vec{m}$ [via the $\mathcal{A}$ field in Eq.~\eqref{eq:eom_den}].  This yields $m \propto \theta_{\mathrm{sH}} \mathcal{A} J$.  Secondly, the DMC couples $\vec{J}$ and $\vec{m}$ directly in Eq.~\eqref{eq:eom_den}, which gives $m \propto \kappa J$.

In most spintronic devices, the spin Hall angle is small ($\theta_{\mathrm{sH}} \ll 1$)~\cite{sinova2015spin}. Let us assume that the conversion factors between $\{\vec{J},\boldsymbol{\mathcal{J}},\vec{m}\}$ are all of the same order, $\theta_{\mathrm{sH}}  \sim v_F \tau_c \mathcal{A} \sim v_F \tau_c \kappa_j^a \ll 1$~\cite{chunli2016direct}.  To lowest order in $\theta_{\mathrm{sH}}$, $\mathcal{A}$, and $\kappa_j^a$, and assuming perfect charge screening (i.e., taking $\rho$ to be uniform), we can take $J_i \approx \sigma_D E_i$.  Then we can combine Eqs.~\eqref{eq:eom_den} and \eqref{eq:eom_sc}, and take the steady-state limit,
%
to arrive at the following steady-state diffusion equation:
\begin{align}
D [\nabla^2 m(\vec{r})]^a - \frac{m^a(\vec{r})}{\tau_s}  +\omega_{L} \epsilon^{abc}\mathrm{\hat{B}}^b m^{c}(\vec{r}) \notag \\
=  - \sigma \left( \partial_i  \gamma_{ij}^{a} \partial_j + \kappa_{j}^{a} \partial_j \right) \phi(\vec{r}). \label{eq:diff_steady}
\end{align}
Here, $\phi$ is the electrostatic potential, which obeys Laplace's equation $\partial_{i}^2 \phi(\vec{r}) =0$.  The left side of Eq.~\eqref{eq:diff_steady} describes the transport of $\mathbf{m}$, and the right side describes a spin torque driven by the applied field.  This torque has contributions from both the SHE and the DMC.  Note that the Rashba-Edelstein effect does not contribute to the torque, to leading order in the conversion factors (i.e.~$\{ \mathcal{A},\theta_{\mathrm{sH}}, \kappa_j^a \}$) between $\{\vec{J},\boldsymbol{\mathcal{J}},\vec{m}\}$.

\textit{Nonlocal Resistance---}We now solve Eq.~\eqref{eq:diff_steady} for a concrete example consisting of adatom-functionalized graphene in a H-bar geometry.  The single-layer graphene sheet is decorated with non-magnetic impurities that are symmetric under rotation, time-reversal and in-plane reflection~\cite{chunli2016direct}.  By symmetry, the only non-vanishing components of the gauge field and the DMC parameter are $\mathcal{A}_{y}^{x}= -\mathcal{A}_{y}^{x} = \ell_{R}^{-1}$ and $\kappa^{x}_{y}=-\kappa^{y}_{x}=\ell_{\mathrm{DMC}}^{-1}$, respectively. Here, $\ell_{R}$ is a  length scale associated with the coupling between  $\boldsymbol{\mathcal{J}}$ and  $\vec{m}$ induced by Rashba SOC, while $\ell_{\mathrm{DMC}}$ is a length scale associated with the coupling between $\vec{J}$ and $\vec{m}$.
%
 %
The model parameters $\{\ell_{R}, \ell_{\mathrm{DMC}}, \theta_{\mathrm{sH}}, \tau_{c}, \tau_{s}\}$ can be calculated \textit{ab initio}, or derived from microscopic scattering models \cite{SM}; in particular, $\ell_{\mathrm{DMC}}$ is assumed to arise from the previously-mentioned anisotropic spin precession scattering mechanism~\cite{chunli2016direct}.

%
%
%


The H-bar device has width $W$ and the distance between the terminals is $L$, as shown in Fig.~\ref{fig:scheme}.  A current $I$ is injected at $x = 0$,  so that the boundary conditions along the upper and lower edges are $J_y\left( x, y= \pm W/2\right)=I \delta(x)$ and $\mathcal{J}_{y}^{a} \left( x, y= \pm W/2\right) =0.$  Solutions for Eq.~\eqref{eq:diff_steady} with these boundary conditions can be obtained via numerical integration \cite{beconcini2016nonlocal,xp2016valley}.  To obtain analytical results, however, we assume that the aspect ratio is large ($L \gg W$) and the width is smaller than the spin diffusion length ($W \leqslant\ell_s$)  .  The latter condition is typically satisfied for micrometer-scale devices in the dilute impurity regime~\cite{Tombros2007,balakrishnan2014giant}.  In that case, the $\mathbf{m}$ field does not relax in the $y$ direction, and Eq.~\eqref{eq:diff_steady} reduces to a 1D problem. In the absence of a magnetic field ($\omega_{L}=0$), we find
\begin{align} 
R_\mathrm{nl}(L)
 &\equiv \big[\phi \left( x=L, y= -W/2\right) - \phi\left( x=0, y=W/2 \right) \big]/I
    \nonumber \\
 &= \frac{W}{\sigma} \Bigg[ \frac{\theta_{\mathrm{sH}}^2}{2}
  \mathrm{Re} \Big( q e^{-q L }\Big)
  -\frac{2}{\ell_{\mathrm{DMC}}^2}
  \mathrm{Re} \Big( \frac{1}{q}\, e^{-q L }\Big) \nonumber\\
  & \quad\qquad + \frac{2\ell_{s}\theta_{\mathrm{sH}}}{\ell_{\mathrm{DMC}}}\;
  \mathrm{Im}\,\Big( q e^{-q L} \Big) \Bigg], \label{eq:R_nl_A}
\end{align}
where $q = \ell_s^{-1}\sqrt{1+ 4i \ell_s / \ell_{R}}$, and $\ell_s = \sqrt{D\tau_{s}}$ is the spin diffusion length. When $\ell_s \ll \ell_\mathrm{R}$, the $\ell_\mathrm{R}$ length scale drops out of $R_{\mathrm{nl}}$, which then simply decays exponentially with distance \cite{abanin2009nonlocal}:
\begin{equation}\label{eq:Rnl}
\lim_{ \ell_s /  \ell_{R} \rightarrow 0} R_\mathrm{nl}(L) =
\left(
\frac{\theta_{\mathrm{sH}}^{2}}{2}-\frac{ 2 \ell_s^2}{ \ell_{\mathrm{DMC}}^2 }\right) \frac{W e^{-L/ \ell_{s}}}{\sigma \ell_{s}}.
\end{equation}
However, there is something interesting about the terms inside the parentheses in Eq.~\eqref{eq:Rnl}.  The SHE contributes \textit{positively} to $R_{\mathrm{nl}}$, whereas the DMC contribution is \textit{negative}.  These signs are governed by the Onsager reciprocity principle: the SHE couples $\mathbf{J}$ with $\boldsymbol{\mathcal{J}}$, which have opposite parities under time-reversal, whereas the DMC couples $\mathbf{J}$ and $\mathbf{m}$, which have the same parity under time-reversal.  Note that Eq.~\eqref{eq:Rnl} reduces to the result of Abanin \textit{et~al.}~when $\ell_{\mathrm{DMC}}^{-1}\rightarrow 0$ \cite{abanin2009nonlocal}.

When $\ell_s \lesssim \ell_\mathrm{R}$, Eq.~\eqref{eq:R_nl_A} implies that $R_{\mathrm{nl}}(L)$ is an oscillatory decaying function of $L$, as shown in  Fig.~\ref{fig:Rnl}(a).  The oscillation occurs even in the absence of an applied magnetic field, and can be attributed to spin precession induced by Rashba SOC.  Mathematically, it arises from the covariant derivative, via terms like $[\nabla^2 m]^a \propto \epsilon^{abc}\mathcal{A}_i^b \partial_i m^c$ in Eq.~\eqref{eq:diff_steady}.  For $L \gtrsim \ell_R$, this produces a sign change in $R_{\mathrm{nl}}$, which needs to be distinguished from the negative-$R_{\mathrm{nl}}$ feature discussed in the previous paragraph.  That can be done by checking different $L$ (i.e.~different distances between injecting and measuring terminals).

The model parameters all have an implicit dependence on the chemical potential $\mu$, which can be extracted from the microscopic scattering model described in Ref.~\onlinecite{chunli2016direct}.  The resulting plot of $R_{\mathrm{nl}}$ versus $\mu$ is shown in  Fig.~\ref{fig:Rnl}(b).  We find that $R_{\mathrm{NL}} < 0$ when DMC dominates over the SHE (induced by skew scattering), and $R_{\mathrm{NL}} > 0$ in the opposite case, in agreement with Eq.~(\ref{eq:Rnl}). The peaks in $R_{\mathrm{nl}}$ result from zero-temperature scattering resonances of the impurities \cite{ferreira2014extrinsic}; for finite temperatures and different types of  scattering impurities, the resonant peaks will be less pronounced and $R_{\mathrm{nl}}$ near the Dirac point ($\mu =0$) will be lifted from zero.

\textit{Anomalous Hanle Precession---}We now discuss the effect of an applied magnetic field on $R_{\mathrm{nl}}$.  The magnetic field is usually applied in the 2D plane ($\vec{\hat{B}}=\mathbf{\hat{B}}_{\parallel}$), so that $R_{\mathrm{nl}}$ does not receive any contribution from the conventional Hall effect. Assuming the magnetic field is applied in the direction parallel to the electric field (as in previous experiments \cite{balakrishnan2014giant,kaverzin2015electron,balakrishnan_colossal}), the nonlocal resistance becomes
\begin{align} \label{eq:R_nl_B}
\mathcal{R}_\mathrm{nl}(L) &= \frac{W}{\sigma} \Bigg[ \frac{  \theta_{\mathrm{sH}}^2}{2}
  \mathrm{Re}\left( q e^{-q L }\right)
  -\frac{2}{\ell_{\mathrm{DMC}}^2 }
  \mathrm{Re} \left( \frac{1}{q} e^{-q L }\right) \nonumber \\
  & \quad\qquad + \frac{2\ell_{s} \theta_{\mathrm{sH}} }{ \ell_{\mathrm{DMC}}} \left(\frac{4\ell_{s} / 
    \ell_{R}}{\xi}\right)
  \mathrm{Im} \left( q e^{-q L} \right) \nonumber \\
  & \quad\qquad -\frac{2\theta_{\mathrm{sH}} }{ \ell_{\mathrm{DMC}}} \left(\frac{ \omega_{L} \tau_{s} }{\xi}\right)
  \mathrm{Im} \left(  e^{-q L} \right) \Bigg],
\end{align}
where
\begin{equation}
  q= \frac{1}{\ell_s} \sqrt{ 1 + i \xi },\;\;
  \xi=\sqrt{\left(\omega_{L}\tau_{s}\right)^{2}+\left(4 \ell_{s} / \ell_{R} \right)^{2}}.
\end{equation}
As before, $\omega_{L}$ is the Larmor precession frequency of the applied magnetic field.  Note that if either $\ell^{-1}_{\mathrm{DMC}} = 0$ or $\theta_{\mathrm{sH}} = 0$, then $\mathcal{R}_{\mathrm{nl}}$ is even in $\omega_{L}$ (i.e., symmetric under a reversal in the magnetic field direction \cite{abanin2009nonlocal}).  But if $\ell^{-1}_{\mathrm{DMC}}$ and $\theta_{\mathrm{sH}}$ are both non-negligible, $\mathcal{R}_{\mathrm{nl}}$ will be asymmetric under magnetic field reversal.

In Fig.~\ref{fig:Rnl}(c), we plot $\mathcal{R}_{\mathrm{nl}}$ versus $\mu$ using SOC parameters from a microscopic scattering model \cite{chunli2016direct}.  The oscillation period of $\mathcal{R}_{\mathrm{nl}}$ increases away from the Dirac point, consistent with experimental observations \cite{balakrishnan2014giant}. In the SHE dominated regime ($\ell_{R}^{-1},\ell^{-1}_{\mathrm{DMC}} \rightarrow 0$),  we expand the first term of Eq.~\eqref{eq:R_nl_B} in the strong magnetic field limit ($\omega_L \tau_s \gg 1$), and find that the sign change of $\mathcal{R}_{\mathrm{nl}}$ occurs when $\omega_L \tau_s \sim ( \ell_s/|x|)^2$.  The spin relaxation length $\ell_s= \sqrt{D\tau_s}=v_F \sqrt{2^{-1}\tau_c \tau_s}$ depends on both the elastic scattering time $\tau_c$ and spin relaxation time $\tau_s$, so the critical magnetic field to observe Hanle precession is $B_{c}=(\hbar v_{F}^{2}/g \mu_{B} |x|^2) \, \tau_{c}(\mu) $, proportional to the elastic scattering time (hence mobility). For $\omega_L \tau_s \gg 1$, moderately-doped graphene with chemical potential $\epsilon_{F}=0.1$eV, a high mobility sample \cite{balakrishnan2014giant} ($\mu = 10,000\mathrm{cm}^2/\mathrm{Vs}$) with elastic scattering rate $\tau_{c}=10^{-13}$s, we find a critical magnetic field of $B_c \sim 9 \,\mathrm{T}$ at distance $L=1\mu$m. The weak $\tau_s$ dependence on $B_c$ persists even in moderate magnetic fields ($\omega_L \tau_s \sim 1$)~\cite{SM}.  The elastic scattering time is minimum near the Dirac point (charge neutrality point) \cite{monteverde2010transport}, which explains why the oscillation observed in Ref.~\cite{balakrishnan2014giant} is more pronounced near the Dirac point. Our findings suggest that the discrepancies between recent nonlocal transport experiments on graphene \cite{balakrishnan2014giant,neutral2015wang,kaverzin2015electron} are due to differences in electron mobility.

\begin{figure}
\includegraphics[width=0.45\textwidth]{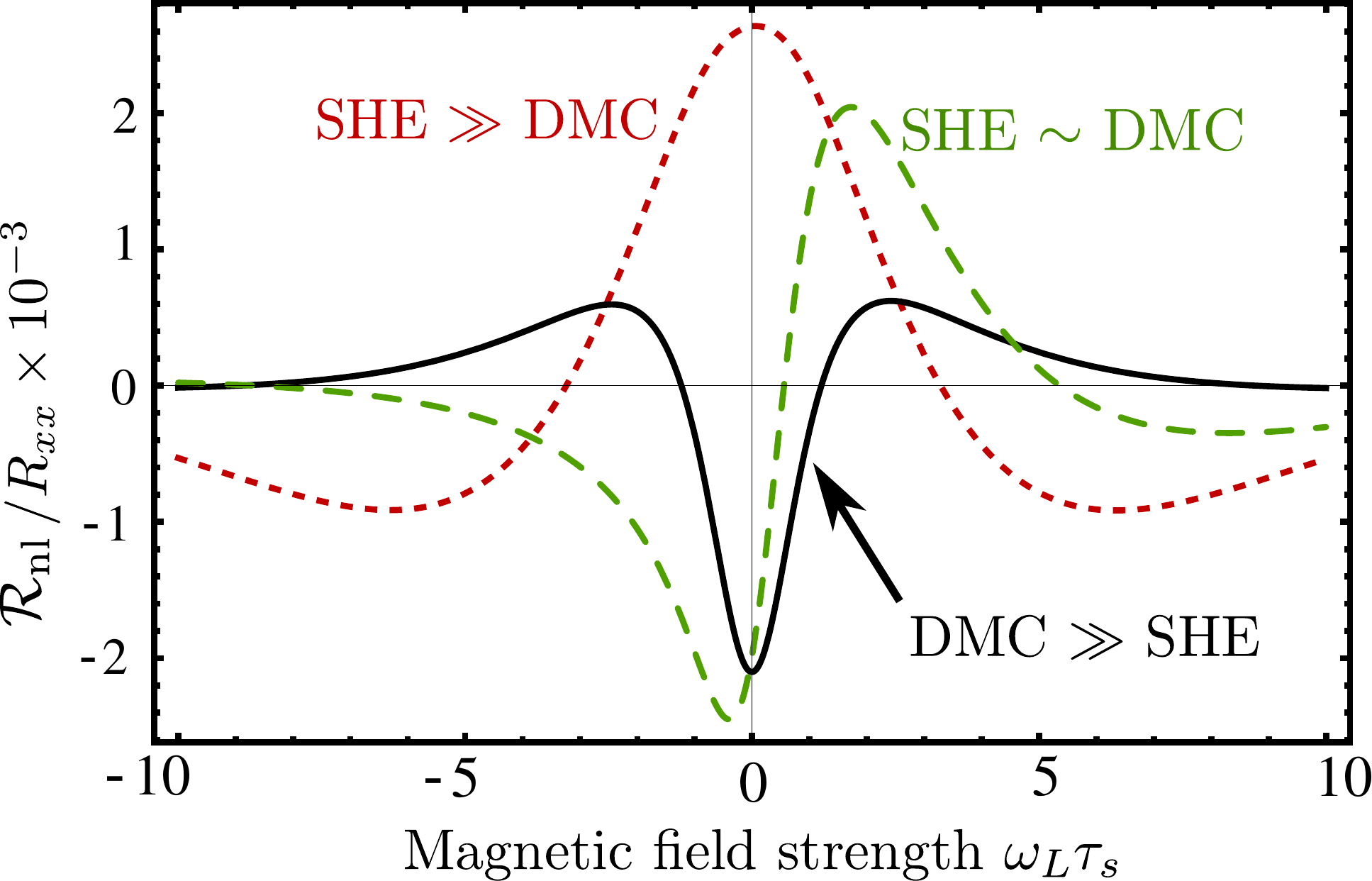}
\caption{Anomalous Hanle precession. The two main spin-charge conversion mechanisms are the spin Hall effect (SHE) and current-induced spin polarization. The latter is characterized by the direct magnetoelectric coupling (DMC) length $\ell_{\mathrm{DMC}}$ \cite{chunli2016direct}. The Hanle precession behaves very different in different limits and can serve as a guide to identify the dominant spin-charge conversion mechanism.
The parameters used in the plots are $\ell_{s}=10^{-6}$m \cite{balakrishnan2014giant} and  $\ell_{R}=10^{-5}$m. For SHE $\gg$ DMC scattering, $\theta_{\mathrm{sH}} =0.2$ and $\ell_s / \ell_{\mathrm{DMC}} =10^{-3}$;  for DMC  $\gg$ SHE, $\theta_{\mathrm{sH}} =10^{-3}$; and $\ell_s/ \ell_{\mathrm{DMC}} =0.1$ and for SHE $\sim$ DMC scattering , $\theta_{\mathrm{sH}}=\ell_s / \ell_{\mathrm{DMC}} =0.1$.
}\label{fig:Rnl_pheno}
\end{figure}

Eq.~\eqref{eq:R_nl_B} can also be regarded as a phenomenological equation  applicable to any 2D metallic system with rotation, time-reversal and in-plane reflection symmetry. The phenomenological parameters $\{\ell_{R},\ell_{\mathrm{DMC}},\theta_{\mathrm{sH}},\tau_c, \tau_s\}$ can arise from different microscopic models. For example, in a 2D electron gas, $\ell_{R}$ can be controlled via the asymmetric confining potential. Results similar to those shown in Fig.~\ref{fig:Rnl_pheno} will then be obtained. In situations where the $\mathbf{J}$-$\mathcal{J}$ coupling (e.g.~SHE) dominates over  DMC, $\mathcal{R}_{\mathrm{nl}}$ will oscillate away from a positive value; in the opposite limit, $\mathcal{R}_{\mathrm{nl}}$ will oscillate away from a negative value, and the oscillation will occur at smaller magnetic fields. When the couplings are of the same order, the Hanle precession curve will be highly assymetric under a sign change of the magnetic field. This may serve as a helpful guide for identifying the spin-charge conversion mechanisms in 2D spintronic materials.
At finite temperature, the nonlocal resistance will be modified by both phonon enhanced SOC \cite{PhysRevB.92.035421,PhysRevB.86.245411} and phonon-induced skew-scattering \cite{PhysRevLett.115.076602}. However, the competition between the two should not dramatically modify the anomalous nonlocal resistance features proposed in the letter. The investigation of the microscopic origins of the temperature dependence of DMC and SHE is beyond the scope of this work.

\textit{Summary--} We have discussed two anomalous features of nonlocal resistance, induced by the interplay between diffusion, spin coherent dynamics, and the direct coupling between charge current and spin polarization. The presence of direct magnetoelectric coupling can give rise to negative nonlocal resistance, even without a magnetic field; when a magnetic field is applied, it gives rise to anomalous Hanle precession.  In general, DMC exists in 2D metallic systems lacking spatial inversion symme- try, such as 2D electron gases confined in semiconductor quantum wells \cite{chunli2017spin}. These features can be used as an experimental probe for the relative strengths of different spin-charge conversion mechanisms in the sample.  Our results shed light on recent nonlocal transport experiments on graphene~\cite{balakrishnan2014giant,neutral2015wang,kaverzin2015electron}.  They may also help explain recent experimental observations of negative nonlocal resistance in gold \cite{mihajilov2009prl}, where a negative nonlocal resistance was reported but interpreted in terms of a ballistic transport model. Note that, unlike the nonlocal spin valve experiment, the nonlocal resistance discussed in this letter does not involve spin injection and detection from spin polarized contacts.

The authors acknowledge support by the Ministry of Science and Technology (Taiwan) under contract No.~NSC~102-2112-M-007-024-MY5, by Taiwan's National Center of Theoretical Sciences (NCTS), by the Singapore National Research Foundation grant No.~NRFF2012-02, and by the Singapore Ministry of Education Academic Research Fund Tier 2 Grant No. MOE2015-T2-2-008. 
\appendix

\bibliography{reference}

\pagebreak
\widetext
\begin{center}
\textbf{\large Supplementary Materials: Anomalous nonlocal resistance and spin-charge conversion mechanisms in 2D metals}
\end{center}
\setcounter{equation}{0}
\setcounter{figure}{0}
\setcounter{table}{0}
\setcounter{page}{1}
\makeatletter
\renewcommand{\theequation}{S\arabic{equation}}
\renewcommand{\thefigure}{S\arabic{figure}}
\renewcommand{\bibnumfmt}[1]{[S#1]}
\renewcommand{\citenumfont}[1]{S#1}

\section{Drift-diffusion Equations of Spin and Charge}
In this section, we discuss in detail the derivation of drift-diffusion equations from the quantum Boltzmann equation (QBE). The spatially uniform QBE is derived in Ref.~[34]. In order to incorporate diffusion, the QBE is generalized to include the diffusion term as follow:
\begin{align}
\label{eq:QBE}
\partial_t \delta n_k(\vec{r},t) + \vec{v}_{k}\cdot \nabla_{r} \delta n_k(\vec{r},t) + \frac{i}{\hbar} \gamma \left[ \delta n_k(\vec{r},t) , \boldsymbol{s}\cdot \boldsymbol{\mathcal{H}}(\vec{r},t)\right] 
+ e \boldsymbol{E}(\vec{r},t)\cdot \frac{\boldsymbol{\nabla}_{k} n^0_k}{\hbar} =
\mathcal{I}\left[ \delta n_k (\vec{r},t)\right].
\end{align}
Here $n_{k}(\vec{r},t)=n_k^{0}+ \delta n_{k}(\vec{r},t)$ is the distribution function in spin space. $n_k^{0}$ is the equilibrium distribution function
while $\delta n_{k}(\vec{r},t)$ is the out-of-equilibrium distribution reacts to the applied electric and magnetic fields, $\boldsymbol{E}(\vec{r},t)$ and $\boldsymbol{\mathcal{H}}(\vec{r},t)$. Here, $\boldsymbol{s} = \frac{\hbar}{2} \boldsymbol{\sigma}$ is the electron spin operator ($\boldsymbol{\sigma} = (\sigma^x, \sigma^y, \sigma^z)$ are the Pauli matrices) and $\gamma$ is the gyromagnetic ratio. The external electric and magnetic fields are assumed to vary slowly compared to the Fermi scale (i.e. $q \ll k_F$ and $\omega \ll \omega_F$ where $q$ ($\omega$) is the wavelength (frequency) of the external field and $k_F $ ($\omega_F$) is the Fermi momentum (frequency). We neglect the correction to the velocity operator arising from side-jump mechanism and take $\vec{v}_k=\hbar k/m^*$, where $m^*$ is the quasiparticle mass.
To leading order in impurity density $n_{\mathrm{imp}}$, the collision integral is given by the following:
\begin{align}
\label{eq:coll-int}
\mathcal{I}[\delta n_{k}(\vec{r},t)] = \frac{ i}{\hbar}  [\delta n_{k}(\vec{r},t) ,\Sigma^R_{k} ] +\frac{2\pi n_{\mathrm{imp}}}{\hbar} \sum_{\boldsymbol{p}} \delta(\epsilon_{k}-\epsilon_{p})  
 \times \left( \mathcal{T}^{+}_{kp} \delta n_{p}(\vec{r},t) \mathcal{T}^{-}_{pk}-
\frac{\mathcal{T}^{+}_{kp}\mathcal{T}^{-}_{pk}\delta n_{k}(\vec{r},t) + 	\delta n_{k}(\vec{r},t) \mathcal{T}^{+}_{kp} \mathcal{T}^{-}_{pk} }{2} \right).
\end{align}
\noindent
Here $\mathcal{T}^{+}_{kp}\equiv\langle k | \mathcal{T}(\epsilon_k + i0^+)| p \rangle$ ($\mathcal{T}_{pk}^{-}$) is the retarded (advanced) on-shell T-matrix of a single impurity located at the origin. $|k\rangle$ and $
|p\rangle$ are the Bloch eigenstates of the pristine single particle Hamiltonian. Given the T-matrix, the collision integral can be evaluated using the following \textit{ansatz} [34]:
\begin{equation} \label{eq:ansatz}
n_{k}(\vec{r},t)=n^0_k  +\delta n_{k}(\vec{r},t) \\= f_{\mathrm{FD}}\left[\epsilon_k - \mu(\vec{r},t)- \hbar \boldsymbol{k} \cdot \boldsymbol{v}_{c}(\vec{r},t) - ((\hbar\boldsymbol{k} \cdot \boldsymbol{v}_{s}(\vec{r},t))\boldsymbol{\hat{n}}_{1} + h_0 \boldsymbol{\hat{n}}_{0}(\vec{r},t))  \cdot \boldsymbol{\sigma}\right].
\end{equation}
Here, $n^0_k = f_{\mathrm{FD}}(\epsilon_k)$ is the equilibrium Fermi-Dirac distribution function. Our \textit{ansatz} assumes the out-of-equilibrium system reach a local-instantaneous equilibrium state described by $\mu(\vec{r},t)$,  $\boldsymbol{v}_{c}(\vec{r},t)$, $\boldsymbol{v}_{s}(\vec{r},t)$) and  $h_0(\vec{r},t)$ whose dynamics are very slow in the long wavelength limit. $\mu(\vec{r},t)$ is the local-instantaneous chemical potential, $\boldsymbol{v}_{c}(\vec{r},t)$ ($\boldsymbol{v}_{s}(\vec{r},t)$) is the drift velocity of the charge (spin) degrees of freedom and $h_0(\vec{r},t)$ is proportional to the magnitude of the magnetization; $\boldsymbol{\hat{n}_0}$ and  $\boldsymbol{\hat{n}_1}$ are the directions of magnetization and spin current polarization respectively. The quantities of interest are the charge density $\rho(\vec{r},t)$, the magnetization (i.e. non-equilibrium spin polarization), $\boldsymbol{M}(\vec{r},t) = (M^x, M^y, M^z)$, the charge current (flow) density, $\boldsymbol{J}(\vec{r},t) = (J_x, J_y)$, and  the spin current (flow) density $\boldsymbol{\mathcal{J}}^{a}(\vec{r},t) = (\mathcal{J}_x^a, \mathcal{J}_y^a)$  (where $a=x,y,z$ is the spin orientation).
 At zero temperature, they are related with the \textit{ansatz} by 
\begin{align} 
\rho(\vec{r},t) =& \frac{1}{2 \Omega}\sum_{k}\mathrm{Tr}[ n_{k}(\vec{r},t)\sigma^{0}]=\bar{\rho}+ N(\mu)\mu(r,t), , \\
m^a (\vec{r},t)=&\frac{1 }{2\Omega} \sum_{k}\mathrm{Tr}[ n_{k}(\vec{r},t)\sigma^{a}] = \hbar  N(\mu) h_{0}(\vec{r},t)   \: (\hat{n}_{0})^a,\\
J_i (\vec{r},t)=& \frac{1}{2\Omega}\sum_{k}(\boldsymbol{v}_{k})_{i} \mathrm{Tr}[ n_{k}(\vec{r},t)\sigma^{0}] =e N(\mu)\epsilon_{F}\frac{(v_{c}(\vec{r},t))_i}{2},\\
\mathcal{J}^{a}_i (\vec{r},t)=&\frac{1}{2\Omega} \sum_{k}(\boldsymbol{v}_{k})_i  \mathrm{Tr}[n_{k}(\vec{r},t)\sigma^{a}] =e  N(\mu)\epsilon_{F}\frac{(v_{s})_{i} (\vec{r},t)  (\hat{n}_{1})^a }{2},
\end{align}
where $\Omega$ is the area of the 2D material, $\boldsymbol{v}_{k} = \hbar v_F (\mathbf{{k}}/k)$ is the group velocity, and $N(\mu)$ is the density of states at Fermi energy. Here $\bar{\rho}=k_F^2/4\pi$ is the average density of electrons. In discussing the drift-diffusion equations, it is useful to work with the convention where charge density and magnetic density are measured in the same units with dimension $L^{-2}$; and charge current density and spin current density are also measured in the same units with dimension $L^{-1}T^{-1}$. This difference in units should not cause any confusion with Ref.~[34]. For graphene, the quantities above should multiply $g_v=2$ to account for the valley degeneracy. To proceed further, we parameterize the T-matix as follow 
\begin{equation}
\mathcal{T}^{+}_{kp}=A_{kp} \: \mathbb{I} + \boldsymbol{B}_{k p} \cdot \boldsymbol{\sigma}.
\end{equation}
Here $A_{kp}$ is the scalar potential and $\boldsymbol{B}_{kp}$ is the ``magnetic field'' in momentum space induce either by magnetic potential and/or spin-orbit coupling potential. The generic parameterization of the QBE in terms of  $A_{kp}$ and $\boldsymbol{B}_{kp}$ are given in Ref.~[34]. We assume here that the impurity potentials are symmetric under in-plane mirror reflection $\mathcal{P}$, time-reversal $\mathcal{T}$  and in-plane rotation $\mathcal{R}$ (in the continuum limit), then the on-shell T-matrix parameters are given by the following (see Appendix of Ref.~[34]):
\begin{align} \label{eq:Tmatrix}
A_{kp}=   a(E, \theta), \;\;
  \boldsymbol{B}_{kp}= \left( \alpha(E, \theta) \sin \left(\frac{\phi}{2}\right)
   \; , \;- \alpha(E, \theta) \cos \left(\frac{\phi}{2}\right), \; b(E, \theta)\right).
\end{align}
Here $E$ is the energy of the incoming electron; $\theta=\theta_k-\theta_p$ is the scattering angle and $\phi=\theta_k+\theta_p$ where $\theta_k \equiv \tan^{-1}(k_y/k_x)$ being the azimuthal angle for vector $k$. Due to the $\mathcal{P},\mathcal{T},\mathcal{R}$ symmetries, the functions $a,b,\alpha$ satisfy the following  properties:
\begin{equation}
 a(E,-\theta)= a(E, \theta) \; , \; b(E,-\theta)=-b(E, \theta)\; , \; \alpha(E,-\theta)=\alpha(E, \theta).
\end{equation}
The odd function $b$ gives precisely the skew-scattering.

To proceed further, we substitute Eq.~\eqref{eq:Tmatrix} and \eqref{eq:ansatz} into Eq.~\eqref{eq:QBE} and arrive at the following closed set of equations:
\begin{align}
\partial_t \rho + \partial_i J_i =&\,  0, \label{eq:r1}
\\
\partial_t  m^a +\partial_i \vec{\mathcal{J}}_{i}^{a}  + \omega_{L}\epsilon^{abc}\hat{ \mathcal{H}}^{b}   m^c =& -\frac{m^a  }{\tau_{s}} + \frac{\epsilon^{a}_{j} J_j }{\ell_{\mathrm{asp}}}  + \epsilon^{abc}\mathcal{A}_{i}^{b} \mathcal{J}_{i}^{c}, \label{eq:r2}
\\
\partial_t  J_i   + \frac{v_F^2}{2} \partial_{i} \rho - \frac{\sigma_{D}}{\tau_{c}} E_{i} =& -\frac{J_i  }{\tau_{c}} - \frac{\epsilon^{a}_{i} m^{a}  }{\ell_{\mathrm{asp}}}+ \alpha	_{\mathrm{sk}}\epsilon_{ij}\delta^{az} \mathcal{J}_{j}^{a} ,
\label{eq:r3}
 \\
\partial_t  \mathcal{J}_{i}^{a}  + \frac{v_F^2}{2} \partial_{i} m^{a}  + \omega_{L}\epsilon^{abc}\hat{ \mathcal{H}}^{b}  \mathcal{J}^c 
=& -\frac{J_i^{a}}{\tau_{c}}  + \alpha_{\mathrm{sk}}\epsilon_{ij}\delta^{az} J_{j}  + \frac{v_F^2}{2}\epsilon^{abc}\mathcal{A}^{b}_{i}m^c. \label{eq:r4}
\end{align}
The left hand side of the equations describe the drift-diffusion response of the system induce by external (electromagnetic) field; here $\sigma_{D}$ is the conductivity and $\omega_{L}$ is the Larmor precession frequency. The right hand side of the equation describes the coupling between different responses (i.e.~$\{ \mathcal{J},J,m \}$) induce by impurities. Note  that charge density  $\rho$  is not coupled with $\{ \mathcal{J},J,m \}$  since it is strictly a conserved quantity. 
The couplings between $\{ \mathcal{J},J,m \}$ are always characterized by a set of three phenomenological parameters whose origin may arise from different microscopic origin depending on the details of the 2D materials. For concreteness, we label the coupling parameters with the skew-scattering rate $\alpha_{\mathrm{sk}}$, the Anisotropic-Spin Precession (ASP) scattering length $\ell_{\mathrm{asp}}$ [34] and the Rashba scattering length $\mathcal{A}^{-1}$. The relaxation of the response are characterized by the elastic scattering time $\tau_c$ and spin relaxation time $\tau_s$. These five parameters $\{ \alpha_{\mathrm{sk}}, \ell_{\mathrm{asp}}, \mathcal{A}, \tau_c,\tau_s \}$ characterized different mechanisms of  spin-charge conversion.
The linear response equation in  Ref.~[34] can be recovered by setting the left hand side of Eqs.~\eqref{eq:r1}--\eqref{eq:r4} except the electric field to zero.

Next, we use the standard approximation and let $\partial_t J=\partial_{t} \mathcal{J}=0$ in the constitutive relationships (Eqs.~\eqref{eq:r3} and \eqref{eq:r4}). This means that the couplings between the responses $\{\rho,J_i,m^a,\mathcal{J}_i^a\}$ in Eqs.~\eqref{eq:r3} and \eqref{eq:r4} are instantaneous. Then, we use the notion of ``covariant'' derivative to simply and arrive at the drift-diffusion equations in the main text:
\begin{align}
&\partial_t \rho + \partial_i J_i =\,  0 ,
\\
&[\nabla_t  m]^a +[\nabla_i  \mathcal{J}_{i}]^{a} = -\frac{m^a  }{\tau_{s}} + \kappa^{a}_j J_j,
\\
 &J_i   = - D \partial_{i} \rho + \sigma_{D} E_{i}  -\tau_{c} \kappa_{i}^{a}m^a + \gamma_{ij}^{a} \mathcal{J}_{j}^{a} ,
 \\
&\mathcal{J}_{i}^{a}   
= - D  [\nabla_{i} m]^{a}  +  \gamma_{ij}^{a}  J_{j}.  \label{eq:r5}
\end{align}
Here $\kappa_{i}^{a}= \ell_{\mathrm{asp}}^{-1} \epsilon^{a}_i$, $\gamma_{ij}^{a}=\theta_{\mathrm{sH}}\epsilon_{ij}\delta^{az}$ and $\theta_{\mathrm{sH}}=\alpha_{\mathrm{sk}}\tau_c$ is the spin Hall angle.
Note that we have neglected a term  proportional  to$\omega_{L} \tau_{c}$ describing the precession of the spin component of the $\mathcal{J}$ in Eq.~\eqref{eq:r5}. This is because the precession of $\mathcal{J}$ governed by $\omega_{L} \tau_{c}$ is normally much smaller than the precession of $m$, which is governed by $\omega_{L} \tau_{s}$. 

%

\section{Microscopic Scattering Model for Adatoms-functionalized Graphene}

The microscopic parameters $\{ \tau_c, \tau_s, \ell_{\mathrm{asp}}, \mathcal{A}, \alpha_{\mathrm{sk}}\}$ can be evaluated from a microscopic scattering models or calculated \textit{ab-initio}  for a particular 2D metals. Using the microscopic scattering model described in  Ref.~[34], the parameters for adatoms functionalized graphene read as follow:
\begin{align}
\alpha_{\mathrm{sk}}&= \frac{\pi n_{\mathrm{imp}} }{  \hbar}N(\mu)\: \mbox{Im}\left( \gamma_{I}\gamma_{0}^{*}\right),  \\
\mathcal{A}_x^{y}&=-\mathcal{A}_x^{y}=\frac{1}{\ell_{R}}=  \frac{  n_{\mathrm{imp}} }{ \hbar v_F} \left( \frac{1}{2}\mathrm{Re}\: \gamma_{R} + \pi N(\mu) \mathrm{Im}\:  (\gamma_{0}+\gamma_{I})\gamma_{R}^* \right), \label{eq:alpha_ext} 
\\
\frac{1}{\ell_{\mathrm{asp}}}&= -\frac{2\pi n_{\text{imp}}}{ \hbar 	v_F} N(\mu)\: \mbox{Re} \gamma_{I}\gamma_{R}^{*}, \label{eq:alpha_cf} \\
\frac{1}{\tau_c}& =\frac{\pi n_{\mathrm{imp}}}{2\hbar}N(\mu)
\left( |\gamma_{0}|^2 + 3| \gamma_{I}|^2 + 4| \gamma_{R}|^2 \right), \\
\frac{1}{\tau_s } &=  \frac{8}{\tau_c }\left( \frac{|\gamma_{I}|^2+ |\gamma_{R}|^{2}}{  |\gamma_{0}|^2 + 3| \gamma_{I}|^2  + 4|\gamma_{R}|^2 } \right),
\end{align}
In this model, all other components of $\mathcal{A}_{i}^{a}=0$ except $\mathcal{A}_x^{y}$ and $\mathcal{A}_x^{y}$. Note that the ASP and Rashba scattering length are related to the ASP and Rashba scattering rates in  Ref.~[34] as follow: $\ell_{\mathrm{asp}}= v_{F}/ \alpha_{\mathrm{asp}}$ and $\ell_{R}= v_F / \alpha_{\mathrm{R}}$. For notational simplicity, we also denote the Elliott-Yafet spin relaxation time $\tau_{\mathrm{EY}}$ in Ref.~[34] simply as the spin relaxation time $\tau_s$.

\begin{figure}[b]
\includegraphics[width=0.45\textwidth]{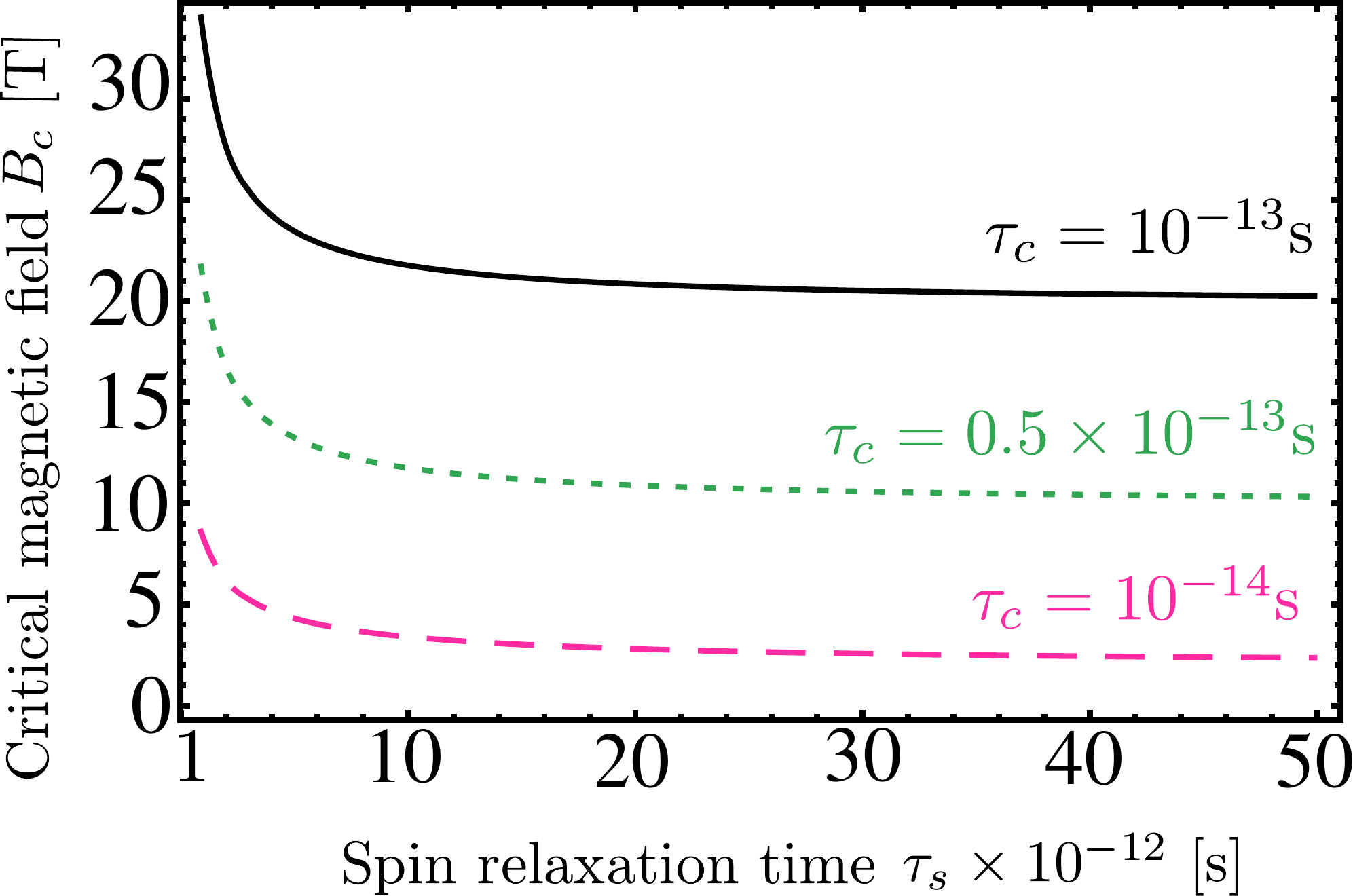}
\caption{The critical magnetic field to observe Hanle precession $B_c$ v.s.~ spin relaxation time $\tau_s$ at various elastic scattering time $\tau_c$. In the spin Hall effect dominated regime, the critical magnetic field depends only on $\tau_c$, $\tau_s$ and the distance between the injection terminals and measurement terminals $L$. $L$ is fixed to $1\mu$m in this plot. The parameter space of $\tau_c$ and $\tau_s$ is chosen such that the spin diffusion length $\ell_s =v_F\sqrt{\tau_c \tau_s/2}\sim L$.
}\label{fig:S_Hanle_pheno}
\end{figure}

The scattering parameters above are functions of $\{ \gamma_0 ,\gamma_I ,\gamma_R \}$. They are complex numbers representing the renormalized potential strength as a function of incident energy, see Ref.~[34] for more information. 
They are related with the bare impurity potential by the following equations:
\begin{align}
\gamma_0 (k)&= \frac{1}{4G_0(k)} \left( \frac{1}{ G_{0}(k) (-\lambda_0+ \lambda_I -2  \lambda_R)+1} +
\frac{1}{G_{0}(k)  (-\lambda_0 + \lambda_I +2\lambda_R)+1}-\frac{2}{G_{0}(k)
   (\lambda_0 +\lambda_I)-1}-4 \right), \\
\gamma_I (k) &= \frac{\lambda_{I}+\lambda_{I}G_{0}(k)(\lambda_{I}-\lambda_{0})-2G_{0}(k)\lambda_{R}^{2}}{\left(1-G_{0}(k)(\lambda_{I}+\lambda_{0})\right)\left(1-G_{0}(k)(\lambda_{0}-\lambda_{I}-2\lambda_{R})\right)\left(1-G_{0}(k)(\lambda_{0}-\lambda_{I}+2\lambda_{R}))\right)} ,\\
\gamma_R (k) &= \frac{\lambda_{R}}{\left(1+G_{0}(k)(\lambda_{I}-\lambda_{0})\right)^{2}-4G_{0}^{2}(k) \lambda_{R}^{2}},
\end{align}
where
\begin{align}
G_0(k)=\text{sign}(E)\frac{k}{2\pi\hbar v_F}\log |kR|-\frac{ik}{4\hbar v_F},
\end{align}
is the Green function at the origin. It is  obtained by imposing a cut-off at  momentum $k^{\prime}\sim R^{-1}$. Here $\lambda_0$, $\lambda_I$ and $\lambda_R$ are the bare scalar potential, Kane-Mele type SOC potential and Rashba type SOC potential strengths.

The parameters used in the Fig.~2 are as follow. In Fig 2a), all conversion factors are set to be $0.1$. For Fig.~2b), for DMC $\gg$ SHE ,$\lambda_0 = 100\mathrm{meV}$,$\lambda_{I} = 22\mathrm{meV}$ and $\lambda_{R} = 25\mathrm{meV}$ while for SHE $\gg$ DMC, $\lambda_0 = 100\mathrm{meV}$,$\lambda_{I} = 25\mathrm{meV}$ and $\lambda_{R} = 1\mathrm{meV}$. Fig.~2c) used the data for SHE$\gg$ DMC and the distance is fixed to be $L=1\mu$m.

Lastly, we discuss the critical magnetic field $B_c$ to observe the Hanle Precession in adatoms-functionalized doped graphene. The observed Hanle precession in Ref.~[8] is quite symmetrical under the sign change $\omega_L \rightarrow -\omega_L$ and this suggests that the spin Hall effect is the dominant spin-charge conversion mechanism. Hence, the nonlocal resistance (Eq.~9 in the main-text) can be approximated by the following formula
\begin{equation}
\lim_{\ell_{\mathrm{DMC}} \rightarrow \infty } \mathcal{R}_{\mathrm{nl}}(L) \rightarrow \frac{W}{\sigma} \frac{  \theta_{\mathrm{sH}}^2}{2}
  \mathrm{Re}\left( q e^{-q L }\right)\label{eq:Rnl_ab}.
\end{equation}
Here $q=\ell_s^{-1}\sqrt{1+ i \omega_L \tau_s}$ and  $\ell_s=\sqrt{D\tau_s}=v_F\sqrt{2^{-1}\tau_c \tau_s}$ is the spin relaxation length which depends on both spin relaxation time $\tau_s$ and elastic scattering time $\tau_c$. This is the same formula derived by  Abanin.~\textit{et.al}. The smallest roots of Eq.~\eqref{eq:Rnl_ab} defines the critical magnetic field $B_c$ to observe Hanle precession. Figure S1.~shows $B_c$ as a function of $\tau_s$ at various $\tau_c$. 
Note that $B_c$  depends more sensitively on the elastic scattering time $\tau_c$ (hence mobility) than on the spin relaxation time $\tau_s$.


\end{document}